\documentclass[%
 pra,
superscriptaddress,
twocolumn,
 amsmath,amssymb,
 aps,
]{revtex4-1}

\usepackage{graphicx}
\usepackage[space]{grffile}
\usepackage{latexsym}
\usepackage{textcomp}
\usepackage{longtable}
\usepackage{multirow,booktabs}
\usepackage{amsfonts,amsmath,amssymb}
\usepackage{natbib}
\usepackage{url}
\usepackage{hyperref}
\hypersetup{colorlinks=false,pdfborder={0 0 0}}
\usepackage[utf8x]{inputenc}
\usepackage[english]{babel}

\usepackage{graphicx}
\usepackage{dcolumn}
\usepackage{bm}
\usepackage{units}
\usepackage{upgreek}
\usepackage{color}
\usepackage{cancel}
\usepackage{hyperref}
\usepackage{braket}

\newcommand{\black}[1]{\textcolor{black}{#1}}

\begin{document}

\title{Classical model for survival resonances close to Talbot time}

\author{Mikkel F. Andersen}
\affiliation{The Dodd-Walls Centre for Photonic and Quantum Technologies, New Zealand}
\affiliation{Department of Physics, University of Otago, Dunedin, New Zealand}

\author{Sandro Wimberger}
\affiliation{Dipartimento di Scienze Matematiche, Fisiche e Informatiche, Universit\`a di Parma, Campus Universitario, Parco area delle Scienze 7/a, 43124 Parma, Italy}
\affiliation{INFN, Sezione di Milano Bicocca, Gruppo Collegato di Parma, Parco area delle Scienze 7/a, 43124 Parma, Italy}

\bibliographystyle{apsrev4-1}

\begin{abstract}
We present a classical approximation for the peaks of survival resonances occurring when diffracting matter waves from absorption potentials. Generally our simplified model describes the absorption-diffraction process around the Talbot time very well. Classical treatments of this process are presently lacking. For purely imaginary potentials, the classical model duplicates quantum mechanical calculations. The classical model allows for simple evolution of phase-space probability densities, which in the limit of the effective Planck's constant going to zero allows for a compact analytical expression of the survival probability as a function of remaining parameters. Our work extends the range of processes that can be described through classical analogues.
\end{abstract}%


\maketitle

\section{Introduction} 

Optical elements based on diffraction and phase dispersion are ubiquitous in modern society, physics and engineering. Electromagnetic waves are used, for instance, to study microscopic material properties for more than a hundred years \cite{Laue1913, Helliwell}. Matter waves of electrons \cite{EM1,EM2}, neutrons \cite{Neutron}, atoms \cite{Schmied2009} or even heavier particles \cite{Virus1, Virus2} are complement tools to investigate material properties or fundamental aspects and limits of quantum physics.

We focus on atomic matter waves diffracted from a light-induced grating. In the limit of far-detuned light from the atom's internal transitions, such a grating provides a conservative potential that is used, e.g., to produce cw optical lattices \cite{Grimm2000} or pulsed diffraction in the Bragg or Raman-Nath regime. The latter is used to implement the atom-optics kicked rotor, a standard model of classical as well as quantum dynamical systems' theory \cite{Casati, Izr1990, Raizen1999, fishmanDL}. A series of recent experiments \cite{Chai2018a,Chai2018b,Chai2020} applied instead an absorption grating to induce pulses on cold rubidium atoms, realizing an atom-optics "killed" rotor (AOKR), implying the loss of atoms due to resonant absorption. The corresponding potential is imaginary describing the dissipative process. Light and matter waves are scattered or diffracted also by such an imaginary potential, an effect that is hard to imagine classically in the sense of either geometric optics or trajectories in phase space.

In this paper, we present a fully classical model for the AOKR that faithfully reproduces the quantum dynamics of atoms close to the Talbot resonances, at which periodic phase revivals occur after single diffraction pulses \cite{Deng1999}. While such a model existed for diffraction from conservative potentials, it is remarkable that it can be amended to describe quantum scattering from purely imaginary potentials which is thought to be a wave effect. 
Please note that while usually dissipation is supposed to induce a quantum-to-classical transitions, in our case, the dissipative grating is usually seen as to induce a purely quantum scattering effect, without a classical analogue. We provide exactly such an analogue classical model whose advantage is its simplicity making it accessible to analytical treatment. Moreover, it provides a classical picture of trajectories in phase space for the iterated dissipative wave diffraction process. \black{In our 
pseudo-classical description the diffraction is modelled by random kicks in momentum space to the atom.}

The remaining of this paper is structured as following. Section \ref{S2} provides the theoretical background for atoms periodically exposed to an optical standing-wave close to resonance with an open transition. Section \ref{S3} introduces the classical model in which absorption results in random momentum kicks. Section \ref{S4} goes in depth with the special case of a resonant standing wave leading to a pure absorption grating, while Section \ref{S5} concludes the paper. 

\section{Theoretical background}
\label{S2} 

We describe a two-level atom with level difference $\Delta E=\hbar \omega_{eg}$ in a standing wave of laser light with frequency $\omega$. This is a good model for rubidium atoms driven by light pulses close to resonance with some chosen internal transition.
The Hamiltonian of the atom in the rotating frame under rotating-wave approximation reads  
\begin{equation}
\hat H = \frac{\hat p^2}{2M} \black{-} (\hbar \Delta + i\hbar \Gamma/2) \ket{e} \bra{e} + \frac{\hbar\Omega}{2} \left( \ket{e} \bra{g}+\ket{g} \bra{e}\right)\cos(k_L \hat x)\,.
 \label{Ham1}
\end{equation} 
$\Omega$ is the Rabi frequency and $\Delta = \omega - \omega_{eg}$  the detuning from the internal transition. $\Gamma$ describes the decay rate of the upper state, and we assume that when the atom in the excited state spontaneously decays it is lost \black{and not further considered in the experiment. For the experimental realisation of the AOKR \cite{Chai2018a,Chai2018b,Chai2020} the ground state has two levels, with only one of them considered here. The dominating decay channel goes into the second level \cite{Chai2018a} that motivates our approximation. Even if the atom would return to the initial ground-state level, it receives a recoil changing its quasimomentum (see below after Eq. \eqref{Uoper2}) and hence its response to the kicking pulses \cite{WGF2003, Clark2021}. This implies that it will be effectively lost from the signal, and only potentially contributing to an incoherent background \cite{Weitz2004, Chai2018a}.}

Neglecting the $\frac{\hat p^2}{2M}$ term for a moment, \black{the Hamiltonian in Eq. \eqref{Ham1}} can easily be diagonalized to give the following eigenvalues:
\begin{widetext}
\begin{eqnarray}
 \lambda_1=-\frac{1}{2}\left(\hbar \Delta + \frac{i \hbar \Gamma}{2}\right) - \frac{1}{2}\sqrt{\left(\hbar \Delta + \frac{i \hbar \Gamma}{2}\right)^2+\hbar^2 \Omega^2 \cos^2 \left( k_lx\right)} \nonumber \\
 \lambda_2=-\frac{1}{2}\left(\hbar \Delta + \frac{i \hbar \Gamma}{2}\right) + \frac{1}{2}\sqrt{\left(\hbar \Delta + \frac{i \hbar \Gamma}{2}\right)^2+\hbar^2 \Omega^2 \cos^2 \left( k_lx\right)}\,.
 \label{eq:eigen}
\end{eqnarray}
\end{widetext}

The probability amplitude to remain in the ground state after a time $t$ is then given by 
\begin{equation}
 G(x,t)=\frac{1}{1-\frac{\lambda_2}{\lambda_1}}\left(\exp \left(-\frac{i}{\hbar}\lambda_2 t \right)-\frac{\lambda_2}{\lambda_1} \exp \left(-\frac{i}{\hbar}\lambda_1 t \right) \right). \label{Gra1}
\end{equation}
that we call the grating operator in what follows. In case of off-resonant driving, i.e. $\Delta \gg \Omega$, we would get 
\begin{equation}
 G(x,t)= \exp \left(- i \frac{\Omega^2}{8\Delta}t \cos (2k_l x)\right)
 \label{Gra2}
\end{equation}

The potential energy $\frac{\hbar\Omega^2}{8\Delta} \cos (2k_l x)$ in this case would be the usual kick potential of the temporal evolution operator of the atom-optics kicked rotor \cite{Raizen1999, Sadgrove2011}. The phase $\phi_d=\frac{\hbar\Omega^2}{8\Delta} t$ is identified as the kicking strength. Since the kicks are supposed to be instantaneous the kicking time $t$ should be short with respect to the inverse of the recoil energy divided by $\hbar$. This is the so called Raman-Nath regime.

Since we are interested in close-to-resonant driving, the grating operator generally does not just give a phase but also an amplitude modulation. Then, the position space representation of the grating operator of Eq. \eqref{Gra1} can be written as
\begin{equation}
 G\left(x\right) = \exp \left(i \Theta\left(x\right) \right) A\left(x\right),
 \label{Gra3}
\end{equation}
where \black{$A\left(x\right) \in [0, 1]$ and $\Theta\left(x\right) \in [0, 2\pi)$ are positive functions determined by Eq. \eqref{Gra1}}, and we assume now that the time $t$ during which the pulse is on is fixed. \black{This time must be short in order to comply with the Raman-Nath condition.} We omit it in the further discussion \black{as it only enters as a parameter in $G$}.

In the standard atom-optic kicked rotor, the time evolution operator for $N$ kicks is given by 
\begin{equation}
 U(0,N) = \left( \exp\left(-\frac{i}{\hbar}\frac{p^2}{2M}\left(\ell\frac{T_T}{2}+\Delta T\right)\right) G \right)^N\,,
 \label{Uoper1}
\end{equation}
with $G$ from Eq. \eqref{Gra2}. Here, $\ell$ is an integer, $\ell\frac{T_T}{2}+\Delta T$ the kicking period with $\Delta T$ the deviation from the nearest integer multiple of half the Talbot time. $\Delta T$ is the small parameter of the classical limit obtained for $\Delta T \to 0$. Now, by substituting the general expression Eq. \eqref{Gra3}, we obtain the one-period time evolution or Floquet operator
\begin{equation}
 U(0,1) = \exp\left(-\frac{i}{\hbar}\frac{p^2}{2M}\left(\ell\frac{T_T}{2}+\Delta T\right)\right) \exp \left( i \Theta \left( x \right) \right) A\left(x\right)   \,.
 \label{Uoper2}
\end{equation}
Following refs. \cite{WGF2004, WGF2003}, we introduce dimensionless variables of the kicked rotor by substituting $\theta=2 k_l x$, $\epsilon=4 \hbar k_l^2 \Delta T /M$, and $J = \epsilon p/2 \hbar k_l$ is the re-scaled momentum. When rewriting Eq. \eqref{Uoper2}, we introduce the fractional part of momentum $\beta=p/(2 \hbar k_l)~\mathrm{(mod~1)}$. Since the grating potential is periodic, the quasimomentum and thereby $\beta \in [0,1)$ is conserved \cite{FGR2002, WGF2003, Sadgrove2011} Eq.~\eqref{Uoper2} then reads
\begin{widetext}
\begin{equation}
 U(0,1) = \exp\left(-\frac{i}{\epsilon}\left(\frac{J^2}{2}+J\pi\ell+2J\pi\ell\beta\right)\right) \exp \left( i \Theta \left( \frac{\theta}{2k_l} \right) \right) A\left(\frac{\theta}{2k_l}\right)   \,.
 \label{Uoper3}
\end{equation}
\end{widetext}

\section{The pseudo-classical model}\label{S3} 

In refs. \cite{FGR2002, WGF2004}, a pseudoclassical model was introduced that successfully describes the atom-optics kicked rotor close to the quantum resonance, where the resonance refers to the Talbot effect causing wave-function revival in-between two kicks. 

We extend this model from \black{ a far-off resonant standing wave} to the close-to or even on-resonant driving of the atom. In the latter case, the potential given by the grating operator Eq. \eqref{Gra3} is completely imaginary corresponding to a pure absorption grating \cite{Schmied2009}. It is remarkable that such a purely classical description is indeed possible for two wave effects, first the diffraction from an absorbing potential, second the rephasing of evolution phases in the external momentum degree of freedom i.e. the Talbot effect.
When we restrict ourselves to a given value of $\beta$, the time evolution induced by Eq. \eqref{Uoper3} can then be approximated by the classical map:
\begin{eqnarray}
 S_{n+1} & = & s\left(\theta_n \right) S_n  
 \label{ecor1} \\
  J_{n+1} &=& J_n + \delta J_n +\epsilon \left. \frac{\partial}{\partial \theta}\right|_{\theta_n} \Theta \left( \frac{\theta}{2k_l}\right)   
  \label{ecor2}\\
 \theta_{n+1} &=& \theta_n+J_{n+1} + \pi \ell + 2 \pi \ell \beta 
 \label{ecor3} \,.
\end{eqnarray}
Here $S_n$ is the overall survival probability of a trajectory after $n$ pulses, $s(\theta)$ the probability that an atom at position $\theta$ will survive a single pulse, and $\delta J_n$ a random momentum kick an atom receives. by the pulse.
Please note that the sign of $\epsilon$ enters the definition of $J$, contrary to the notation used in \cite{FGR2002, WGF2003, WGF2004}. The iteration of this map approximates well the quantum evolution as long as $\epsilon$ remains small. 
\black{For fixed $\epsilon$, the approximation will eventually break down as the number of applied kicks $N$ increases. Hence, for small kick numbers rather large $\epsilon$ are fine, while for large numbers the correspondence will be the better the smaller $\epsilon$ \cite{WGF2003}.}
In contrast to the standard atom-optics kicked rotor, there are two new points to consider in our classical evolution. First, we must iterate also the survival probability of the atoms $S_n$ because of the absorption by the on-resonant grating. Secondly, the random change of momentum $\delta J_n$, derives from the diffraction from the same absorption grating. Both effects are governed by the amplitude function $A\left(\frac{\theta}{2k_l}\right)$ that would be identical to one in the standard atom-optics kicked rotor. Here, however, we obtain for the absorption or mask function the square of the grating amplitude $A$: 
\begin{equation}
 s\left( \theta \right)= A^2\left(\frac{\theta}{2k_l}\right)  \,.
 \label{mask1}
\end{equation}
Considering the case where survival predominantly occurs in the vicinity of the standing wave nodes, then the application of the grating operator broadens the momentum distribution by the Fourier transform of its amplitude $A$. This implies that the distribution from which $\delta J_n$ must be drawn is given by:
\begin{equation}
 \rho\left(\delta J \right)= B\left( \int_0^{2 \pi}{A\left( \theta \right)\exp\left(-i\frac{\delta J}{\epsilon}\theta\right)d\theta} \right)^2, 
 \label{ro}
\end{equation}
where $B$ is a normalization constant.

\begin{figure}
        \centering
	\includegraphics[width=0.9\linewidth]{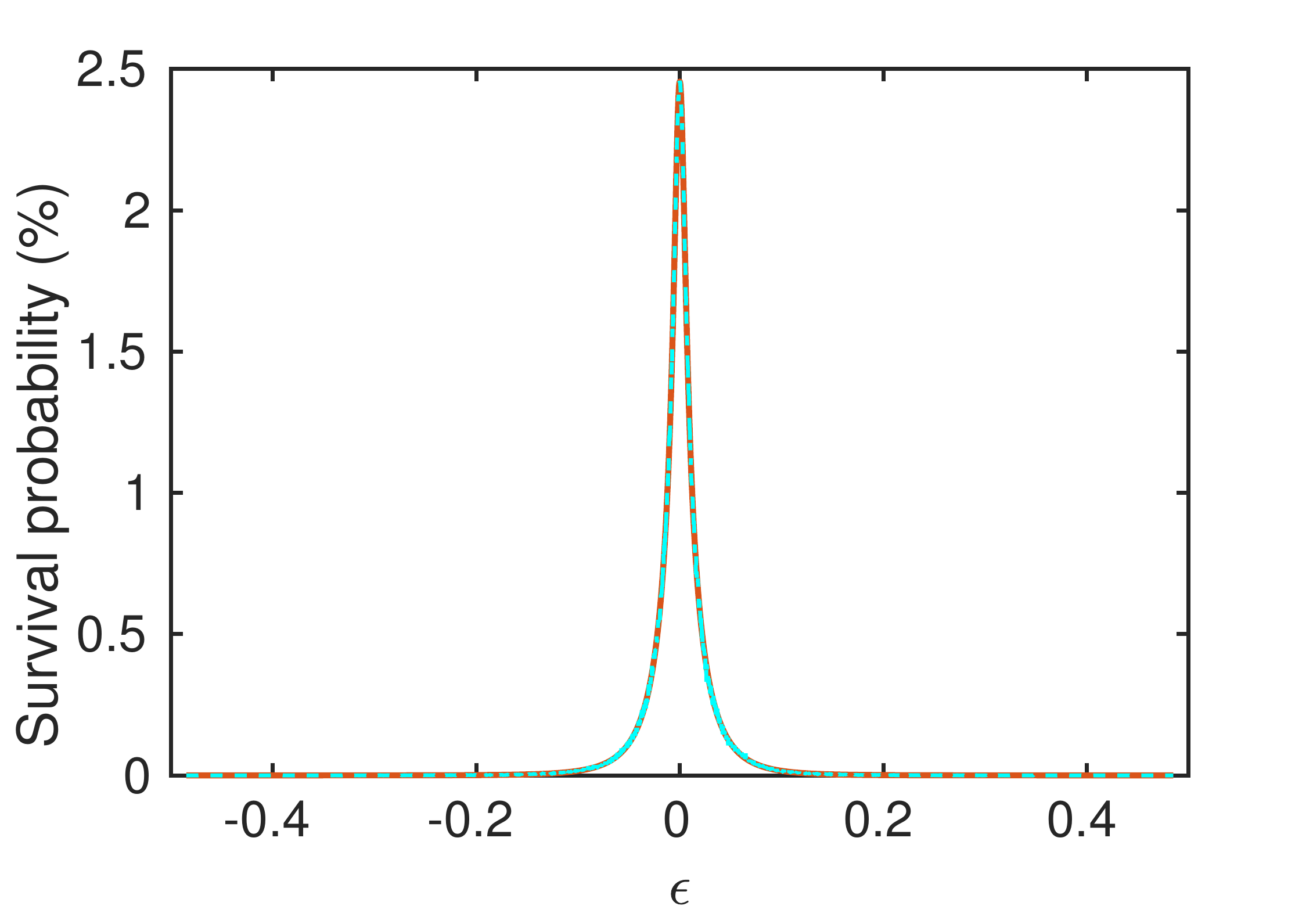}
	\includegraphics[width=0.9\linewidth]{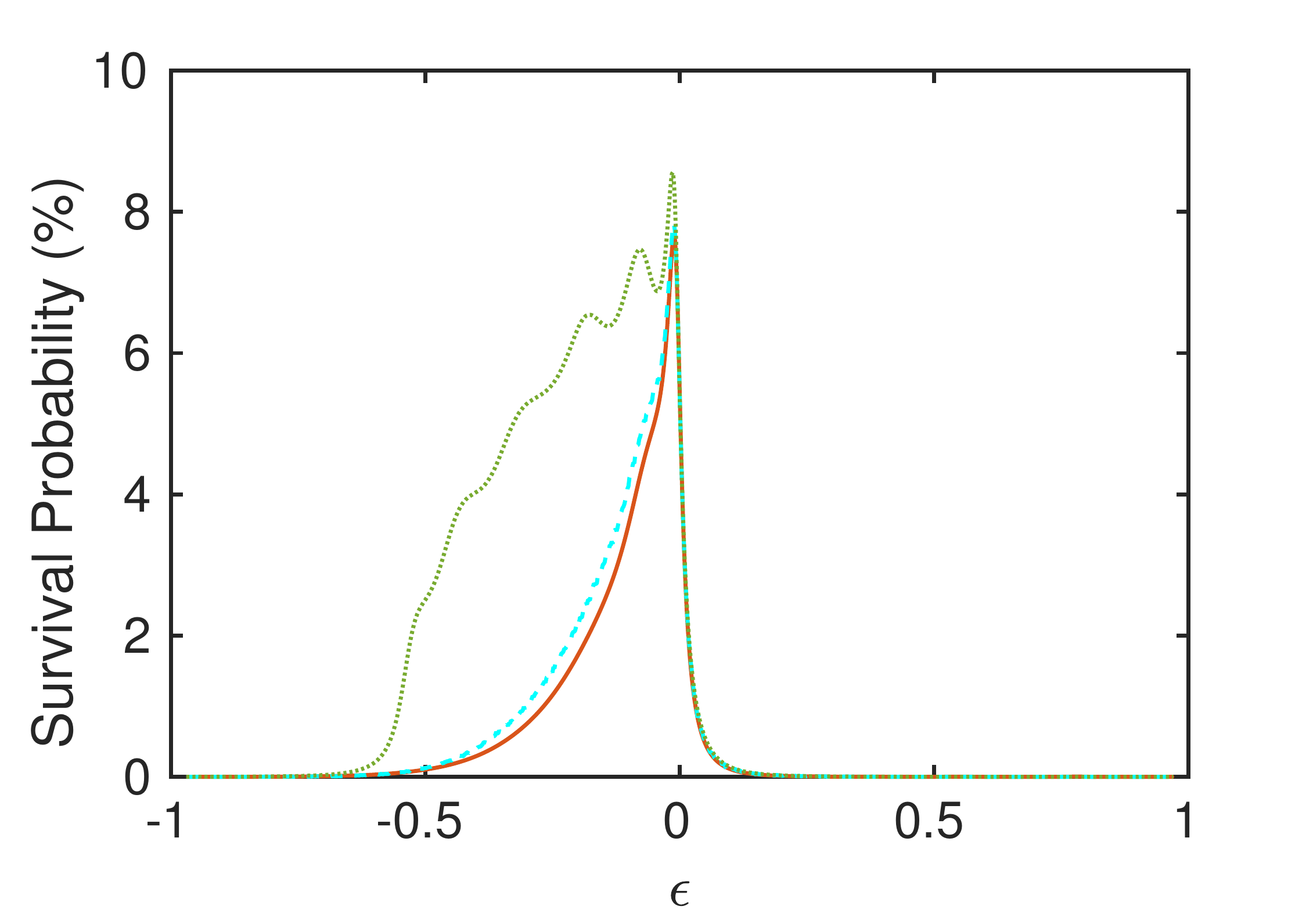}
	\vspace{-4mm}
	\caption{
	(Upper panel) Comparison between $\epsilon$-classical model (dashed cyan line), see Eqs. \eqref{ecor1}-\eqref{ecor3}, and one based on the Floquet operator from Eq. \eqref{Uoper3} (solid red line) for an initial state characterized by $J=\beta=0$ , using resonant standing wave ($\Delta=0$) after $N=7$ kicks.
    (Lower panel) Standing wave off resonance ($\Delta=\Gamma$), but other parameters are the same as in the upper panel. $\epsilon$-classical model (dashed cyan) and quantum evolution (solid red). We observe a good agreement, with minor deviation when $\epsilon>0.1$ is large. The green dotted line is the $\epsilon$-classical result without the random ($\delta J$) kick that indeed plays a crucial role in our classical model of diffraction.
	}
	\label{fig:1}
\end{figure}

Fig. \ref{fig:1} compares the $\epsilon$-classical model based on Eqs. \eqref{ecor1}-\eqref{ecor3}, to the one based on the Floquet operator (Eq. \eqref{Uoper3}). Throughout this work we use the experimentally relevant parameters \cite{Chai2018a,Chai2018b,Chai2020} $k_l=2\pi/780\mathrm{nm}$, $\Gamma=2\pi 6\mathrm{MHz}$, $\Omega=2\Gamma$, $N=7$ kicks, $t=500\mathrm{ns}$, and $M$ the mass of an $^{85}$Rb atom, unless otherwise stated. We see in the figure that the $\epsilon$-classical model captures the quantum dynamics well. Moreover, we see that the random momentum kick $\delta J$ that the particle receives due to the amplitude function remains significant for standing wave detunings of the order of the natural linewidth $\Gamma$. The green dotted line in the lower panel shows the result of omitting $\delta J$, and while we still see resonant behavior the model is no longer in quantitative agreement with the quantum model\black{, in particular for negative $\epsilon$. The reason for the asymmetry is the force term $\epsilon \left. \frac{\partial}{\partial \theta}\right|_{\theta_n} \Theta \left( \frac{\theta}{2k_l}\right)$ in Eq. \eqref{ecor2}, that changes sign with $\epsilon$. For positive $\epsilon$, it will direct atoms away from $\theta=\pi$ where survival is high, leading to the rapid drop in survival when $\epsilon$ increase from zero. On the other hand, for negative $\epsilon$, the force term will direct atoms towards the high survival region making the probability density peak here. However, the addition of a random $\delta J$ will broaden the peak and partially suppress the the enhancement of survival, thereby making the effect of it profound for negative $\epsilon$.   }   

\section{Pure absorptive diffraction}
\label{S4} 

In the case of exactly resonant standing wave light ($\Delta=0$), the above expression simplifies significantly. 
The quantum mechanical exact result for the absorptive diffraction is given, e.g., by Eq. (1) in ref. \cite{Weitz2004}. Using in Eqs. \eqref{eq:eigen}-\eqref{Gra1} a harmonic approximation of the cosine close to the potential minima \cite{Tycho2003} where the atoms have a chance to survive, we obtain as a good approximation:
\begin{equation}
 G(x)=\sum_{j\in \mathbb{Z}}{\exp \left(-\frac{1}{2}\frac{\Omega^2k_l^2}{\Gamma}\left(x-\left(j+\frac{1}{2} \right)\frac{\pi}{k_l}\right)^2 t \right)} 
 \label{sp}
\end{equation}
which is the convolution of a delta-function comb with a Gaussian that has $\sigma_x=\frac{1}{\Omega k_l}\sqrt{\frac{\Gamma}{t}}$.
This directly gives the grating amplitude since there is no phase imprint at all, i.e. $\exp(i\Theta)=1$, implying no $\Theta$-dependent shift in Eq. \eqref{ecor2}. Then the
transmission probability, Eq. \eqref{mask1} for the $n$-kick is
\begin{equation}
 s\left( \theta_n \right)=\exp \left( - \frac{\Omega^2 t}{4 \Gamma}\left(\theta_n - \pi \right)^2\right) \,.
\label{mask2}
\end{equation}
which can be written $s\left( \theta_n \right)=\exp \left( - \frac{\left(\theta_n - \pi \right)^2}{\sigma_\theta^2} \right)$ 
with $\sigma_\theta^2=\frac{4\Gamma}{\Omega^2t}$. Equation \eqref{ro} then yields a distribution from which $\delta J_n$ must be drawn given by:
\begin{equation}
  \rho\left( \delta J \right)= \frac{1}{\sqrt{2\pi \sigma_J^2}}
  \exp \left( - \frac{4 \Gamma}{ \Omega^2 t } \frac{\delta J^2}{\epsilon^2} \right), 
  \label{Jd}
\end{equation}
which is a simple Gaussian with the standard deviation $\sigma_J=\sqrt{\frac{t}{8 \Gamma}}\Omega \epsilon$.
With the explicit results from Eqs. \eqref{mask2} and \eqref{Jd}, the pseudoclassical map can be easily iterated, e.g. for an initial ensemble of atoms, with a given integer and quasimomentum each. Results are shown in Fig. \ref{fig:1}, where we observe a very good correspondence with the full quantum mechanical simulations based on the Floquet operator from Eq. \eqref{Uoper3}. Interestingly, for the resonant standing wave (upper panel) there are no observable deviation at all. 

\subsection{Evolution of phase-space probability densities}

\begin{figure}[t]
\vspace{2mm}\centering
	\includegraphics[width=0.9\linewidth]{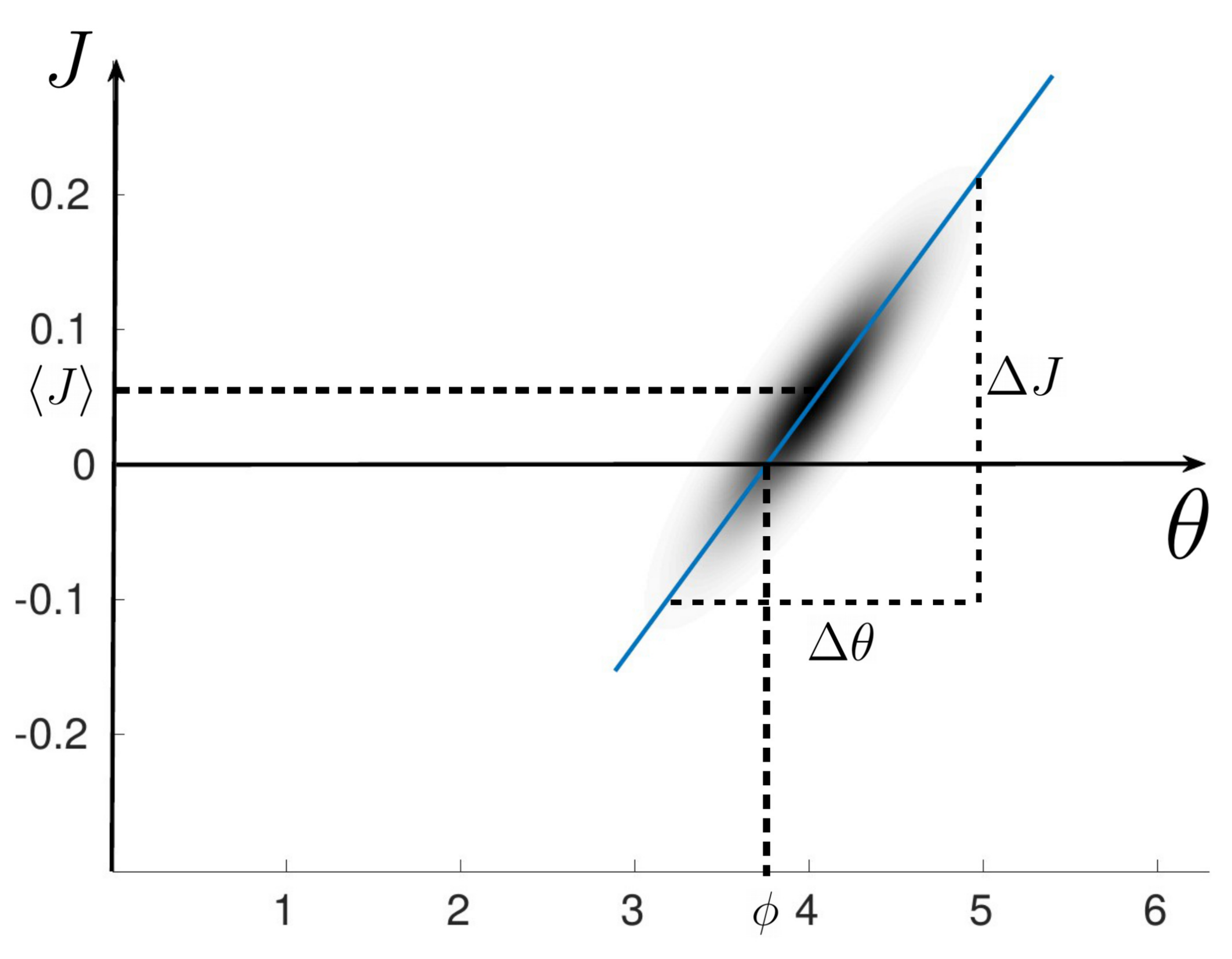}
	\caption{
	The probability density is a Gausian ellipse in phase-space. $\left\langle J \right\rangle$ is the mean of $J$, $\phi$ the intersection of the ellipse's axis with the $\theta$-axis, and $\alpha=\frac{\Delta \theta}{\Delta J}$ is the inverse slope of the ellipse's axis.
	}
	\label{fig:PS}
\end{figure} 

We note that Eqs. \eqref{mask2} and \eqref{Jd} are both Gaussian functions. This means that the $\theta$-$J$ phase-space probability density will always be a (potentially sheared) Gaussian ellipse (see Fig. \ref{fig:PS}). This observation allows for a far more efficient way of computing the survival probability. Generally, the phase-space probability density is characterized by only five numbers $S$, $\left\langle J \right\rangle$, $\alpha$, $\sigma_H$ and $\phi$ (see Fig. \ref{fig:PS}) and can be written:
\begin{eqnarray}
\label{rhow}
\rho_W \left(\theta,J\right)=\frac{S}{\pi \epsilon}  \times \nonumber \\ 
\exp\left(-\sigma_H^2\frac{\left(J-\left\langle J\right\rangle \right)^2}{\epsilon^2}\right)\exp\left(-\frac{\left(\theta-\alpha J -\phi \right)^2}{\sigma_H^2}\right)\,.
\end{eqnarray}
$S$ is the survival probability (the total probability remaining in phase-space), $\left\langle J \right\rangle$ is the mean of $J$ computed with a normalized probability density (that is $S=1$ in Eq. \eqref{rhow}), $\phi$ the intersection of the ellipse's axis with the $\theta$-axis, $\alpha=\frac{\Delta \theta}{\Delta J}$ is the inverse of the slope of the ellipse's axis, and finally $\sigma_H$ determines the width of the probability density along the $\theta$-direction. Please note that the width along the $J$-direction is not a free parameter since it gets dictated by the uncertainty relation originating from $\left[ \theta , J \right]=i\epsilon$.   
 
\begin{figure}
\centering
	\includegraphics[width=0.9\linewidth]{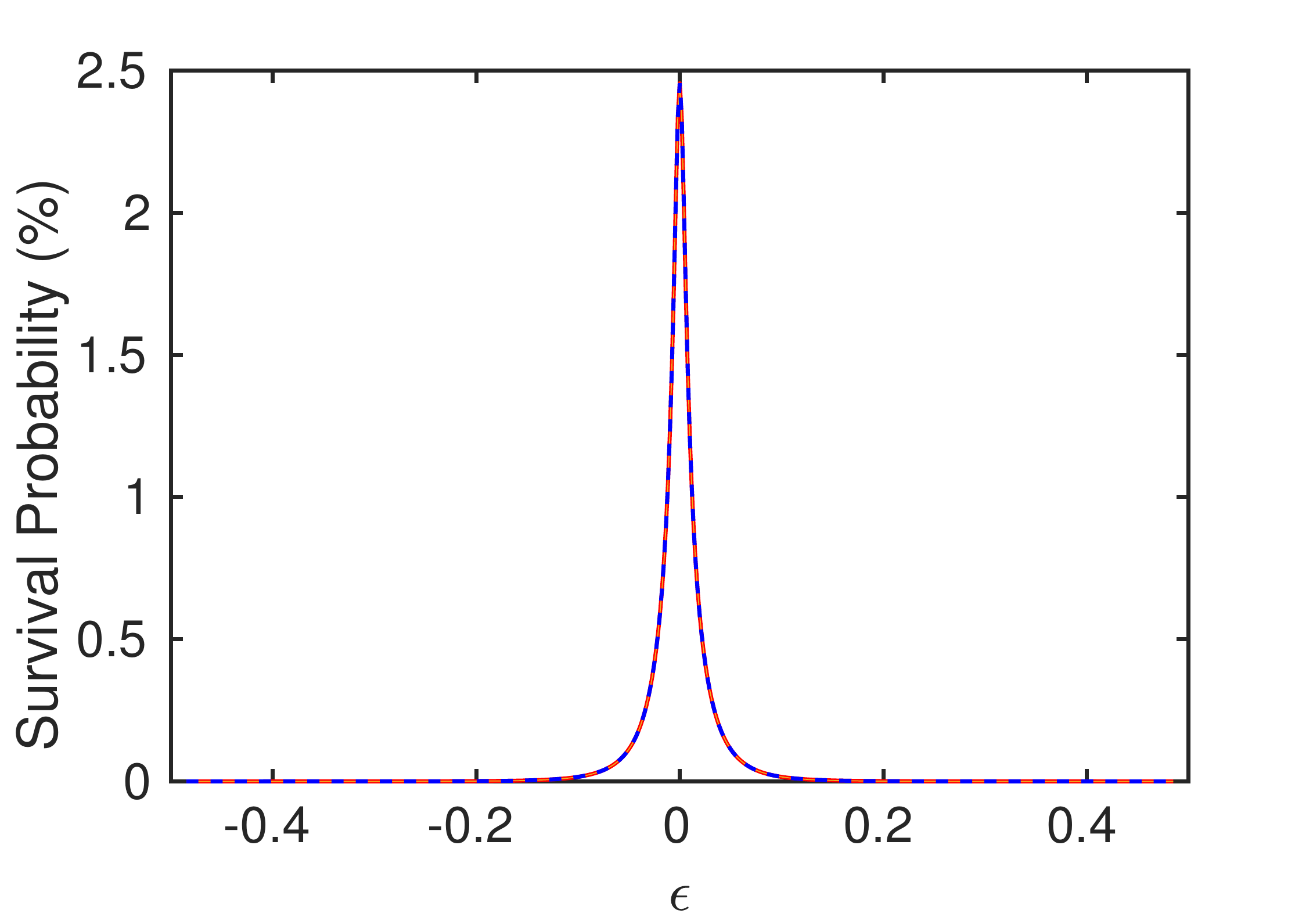}
	\includegraphics[width=0.9\linewidth]{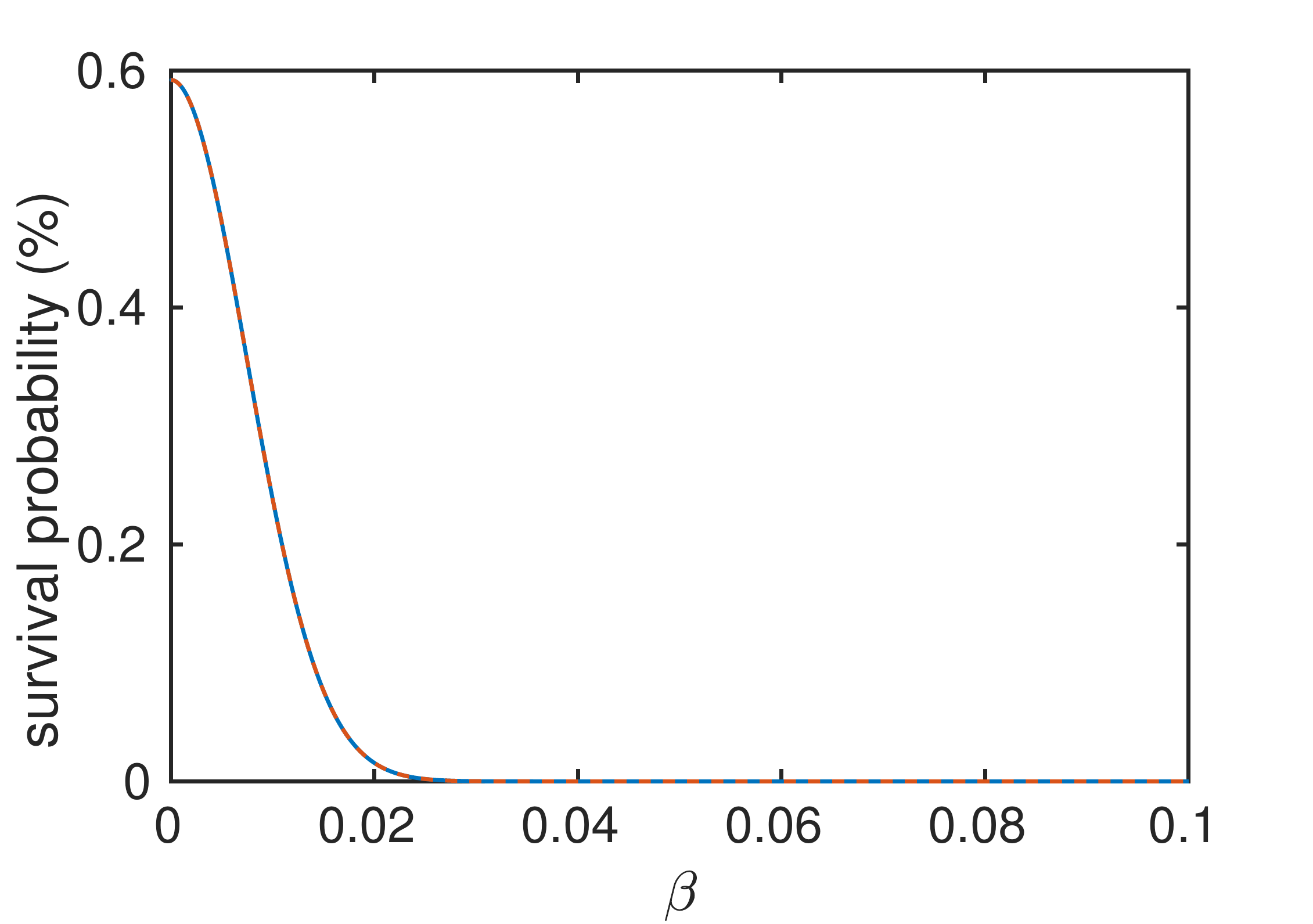}
	\vspace{-4mm}
	\caption{Comparison between recursion model (blue solid) and Eq. \eqref{Uoper3} (red dashed) \black{for $\Omega=2\Gamma$ N=7. Upper panel: Dependence on $\epsilon$ with $\beta=0$. Lower panel: Dependence on $\beta$ with $\epsilon= \left( 10^{-7}~\mu\mathrm{s}\right)\hbar 4 k_l^2/M$.} }\label{recu2}
	\vspace{0mm}
\end{figure}

Rather than evolving individual trajectories using Eqs. \eqref{ecor1}-\eqref{ecor3} we can use the pseudo-classical evolution to propagate the five parameters that characterise the phase-space probability density. This gives recursion relations for these which for the survival probability is:
\begin{widetext}
\begin{equation}
 S_{n+1}= \frac{S_n \sigma_{H,n} \sigma_\theta }{\sqrt{ \left(\sigma_{H,n}^4+\alpha_n^2\epsilon^2 + \sigma_{H,n}^2 \sigma_\theta^2\right)}}\exp \left( \frac{\sigma_{H,n}^2\left(\pi\left(2 \left( \phi_n+\alpha_n \left\langle J \right\rangle_n \right)-\pi \right)-\left( \phi_n + \alpha_n \left\langle J \right\rangle_n\right)^2 \right)}{\sigma_{H,n}^4+\epsilon^2\alpha_n^2+\sigma_{H,n}^2 \sigma_\theta^2}  \right). \label{SN} 
\end{equation}
\end{widetext}
The recursion relations for the remaining four parameters and an outline of how they are found are given in the appendix.
The recursion relations provide an efficient way of computing the survival probability after a given number of pulses. Figure \ref{recu2} compares the survival probability computed in this way with the one computed quantum mechanically based on Eq. \eqref{Uoper3}, and we see the two methods agree excellently with no observable difference between them.  

\black{It may appear surprising that Eq. \eqref{SN} captures the full quantum dynamics for finite $\epsilon$. However, for a resonant standing wave there is no evolution in a potential, and classical and quantum evolution of phase space distributions are identical in free space.  Additionally, the fluctuations from averaging over a finite number of trajectories barely visible in the upper panel of Fig. \ref{fig:1} are no longer present, since Eq. \eqref{SN} evolves the full phase space distribution.}

\subsection{$\epsilon=0$}
In the special case when $\epsilon=0$ it is possible to obtain a compact analytical formula for the survival probability after $N$ standing wave pulses, rather than having to use the recursion relations given by Eqs. \eqref{SN}, \eqref{SH}, \eqref{phir}, \eqref{alp} and \eqref{JM}. In this case $\left\langle J \right\rangle=0$ and the other quantities are independent of $\alpha$, which is undefined. The survival probability after $N$ pulses is:
\begin{eqnarray}
 S_N=\frac{\sigma_\theta}{2\sqrt{\pi}\sqrt{N}} \times \nonumber \\ 
 \exp\left(-\frac{\left( \ell \pi + 2 \ell \pi \beta \right)^2}{12 \sigma_\theta^2}\left(N-1\right)N\left(N+1\right)\right), \label{anabe}
\end{eqnarray}
and expressions for $\phi$ and $\sigma_H$ are in the appendix \footnote{Please note that the angle $\Delta\phi=\ell\pi+2\ell\pi\beta$ should be taken between $-\pi$ and $\pi$ in Eq. \eqref{anabe}.}.

\begin{figure}[b]
\centering
	\includegraphics[width=0.9\linewidth]{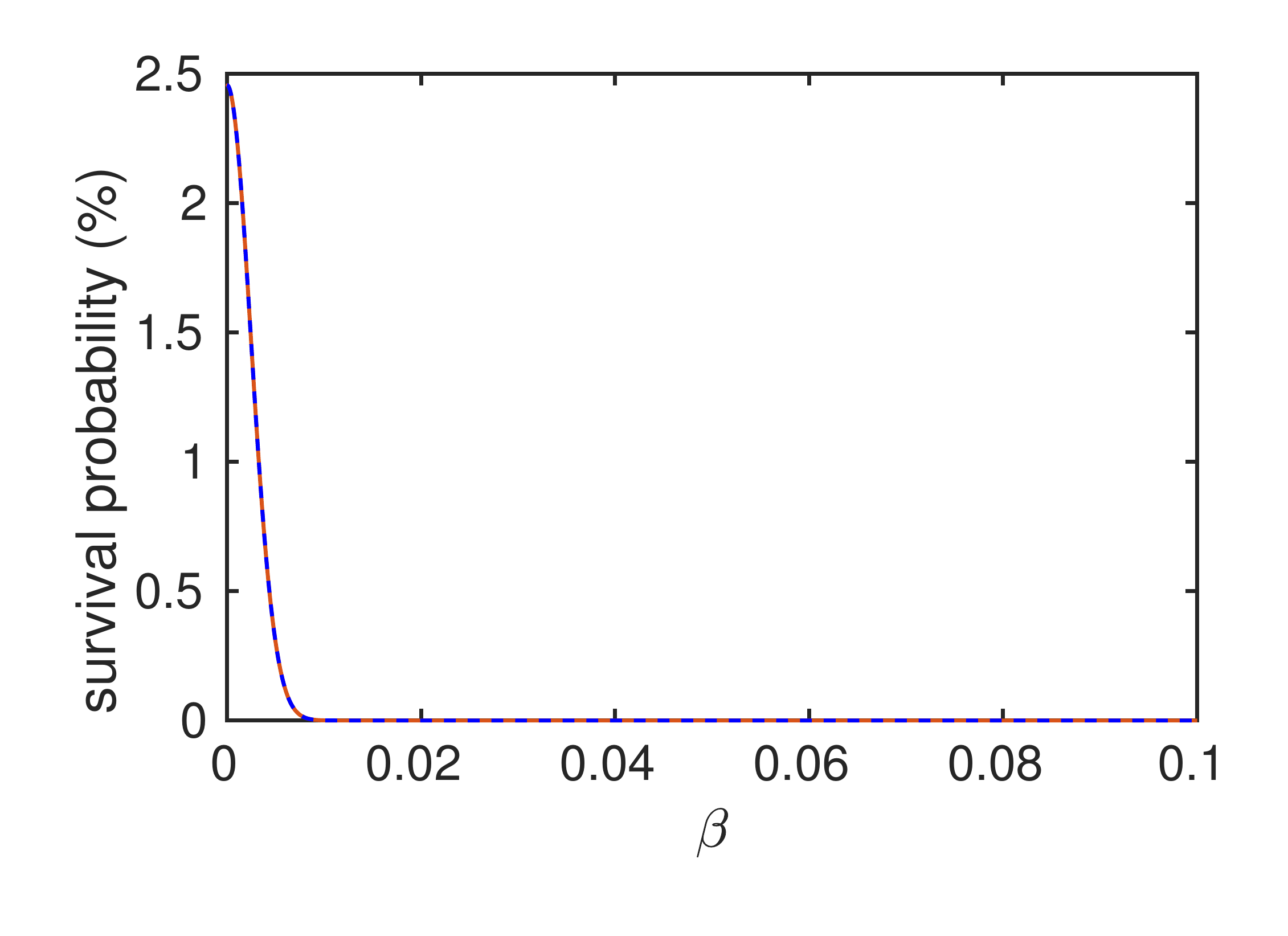}
	\caption{
	The survival probability as a function of $\beta$ for $\epsilon=0$. Red solid line: Calculation based on the Floquet operator from Eq. \eqref{Uoper3}. Blue dashed line: Eq. \eqref{anabe}.
	}
	\label{fig:anabet}
\end{figure} 

Figure \ref{fig:anabet} show a comparison between Eq. \eqref{anabe} and a calculation based on the Floquet operator from Eq. \eqref{Uoper3}. We see that Eq. \eqref{anabe} agrees excellently with the quantum mechanical calculation. This is not surprising since the pseudo-classical model should become strictly correct in the $\epsilon=0$ limit, in analogy to previous work in the off-resonant limit \cite{FGR2002, WGF2004, WGF2003,Sadgrove2011}.

\section{Conclusion}
\label{S5} 

We derived a classical model for the iterated diffraction of matter waves from absorption gratings close to the Talbot revivals. This gives an efficient way of predicting the peak structure of the survival resonances obtained in the experiments \cite{Chai2018a, Chai2018b, Chai2020}. Furthermore, our model offers an intuitive picture of the quantum mechanical diffraction from imaginary potentials based on trajectories in phase space. This way of thinking provides a quantum-to-classical transition for an effect originally treated purely wave-like or quantum mechanical. When the standing wave light is resonant with the atomic transition, leading to pure absorptive diffraction, the classical model agrees remarkably well with quantum mechanical calculations. This situation also allows for a simple analytical expression for the survival probability when $\epsilon=0$ and generally very efficient calculations based on phase-space probability densities.  

\begin{acknowledgments}
SW is very grateful to the Dodd-Walls Centre for Photonic and Quantum Technologies, New Zealand, for a visiting grant that made this work possible.
\end{acknowledgments}

\section*{Appendices}

\begin{widetext}

\section{Recursion map for phase-space probability density}

We propagate a probability density such as Eq. \eqref{rhow} in two steps. First, we compute how it changes due to the standing wave pulse, and then propagate the result using Eq. \eqref{ecor3}.   
Applying the standing wave changes the probability density through multiplying it with Eq. \eqref{mask2} and convolving it along $J$ with Eq. \eqref{Jd}. This follows from Eq. \eqref{ecor1} and \eqref{ecor2}, or simply by calculating the Wigner function after application of the grating operator to the Gaussian probability density. Either way we get:
\begin{eqnarray}
 \rho_{W,a}=
\frac{S_b\sigma_\theta}{\pi\epsilon \sqrt{\left(\sigma_\theta^2+\sigma_{H,b}^2+\frac{\epsilon^2 \alpha_b^2}{\sigma_{H,b}^2}\right)}}\times \nonumber \\
\exp \left(-\frac{\left(\theta-\pi\right)^2}{\sigma_\theta^2}-\frac{\sigma_\theta^2}{\epsilon^2}J^2-\frac{\sigma_{H,b}^2}{\epsilon^2}\left\langle J \right\rangle_b^2-\frac{\left(\theta-\phi\right)^2}{\sigma_{H,b}^2}+\frac{\left(\frac{\sigma_\theta^2}{\epsilon^2}J+\frac{\sigma_{H,b}^2}{\epsilon^2}\left\langle J \right\rangle_b+\frac{\alpha\left(\theta-\phi \right)}{\sigma_{H,b}^2}\right)^2}{\left(\frac{\sigma_\theta^2}{\epsilon^2}+\frac{\sigma_{H,b}^2}{\epsilon^2}+\frac{\alpha_b^2}{\sigma_{H,b}^2}\right)}\right) , \label{neww}
\end{eqnarray}
where the subscript $b$ denotes the value before the grating was applied. Equation \eqref{neww} again describes a Gaussian ellipse in phase-space and noting that Eq. \eqref{rhow} can be written:
\begin{eqnarray}
    \rho_W\left(\theta,J\right) = \frac{S}{\pi \epsilon} \times \nonumber \\ \exp \left( -\frac{1}{\sigma_H^2}\theta^2-\left(\frac{\sigma_H^2}{\epsilon^2}+\frac{\alpha^2}{\sigma_H^2}\right)J^2-\frac{2 \alpha \phi}{\sigma_H^2}J+\frac{2 \sigma_H^2 J_m}{\epsilon^2}J+\frac{2 \theta \alpha}{\sigma_H^2}J-\frac{\sigma_H^2 J_m^2}{\epsilon^2}-\frac{\phi^2}{\sigma_H^2}+\frac{2 \theta \phi}{\sigma_H^2}\right), \label{EQA2}
\end{eqnarray}
we see that the $\theta^2$ term in the exponential function determines $\sigma_H$. So isolating that term in Eq. \eqref{neww} allows us to determine what $\sigma_H$ is after application of the grating in terms of the parameters before. Moreover, by observing Eq. \eqref{ecor3} we see that $\sigma_H$ does not change due to the evolution between standing wave pulses, so we find the following recursion relation for $\sigma_H$ where the subscript $n$ denotes the quantities values before the $n+1$'th standing wave pulse: 
\begin{eqnarray}
 \sigma_{H, n+1}^2=\frac{1}{\frac{1}{\sigma_\theta^2}+\frac{1}{\sigma_{H, n}^2}-\frac{\epsilon^2 \alpha_n^2}{\sigma_{H,n}^2 \left(\sigma_\theta^2\sigma_{H,n}^2+\sigma_{H,n}^4+\epsilon^2\alpha_n^2\right)}}=\\
\frac{\sigma_\theta^2\left(\sigma_{H,n}^2\sigma_\theta^2+\sigma_{H,n}^4+\alpha_n^2 \epsilon^2\right)}{\left(2\sigma_{H,n}^2\sigma_\theta^2+\sigma_\theta^4+\sigma_{H,n}^4+\alpha_n^2 \epsilon^2\right)}. \label{SH}
 \end{eqnarray}

Similarly $\phi$ after application of the standing wave (denoted $\phi_a$) can be determined from the term proportional to $\theta$ but not $J$:
\begin{eqnarray}
 \frac{2 \phi_a}{\sigma_{H,n+1}^2}=\frac{2 \phi_n}{\sigma_{H,n}^2}+\frac{2\pi}{\sigma_\theta^2}+\frac{-\frac{2\epsilon^2 \alpha_n^2  \phi_n}{\sigma_{H,n}^2}+2 \left\langle J\right\rangle_n \sigma_{H,n}^2 \alpha_n}{\sigma_\theta^2\sigma_{H,n}^2+\sigma_{H,n}^4+\epsilon^2\alpha_n^2}\\
 \Updownarrow \nonumber \\
 \phi_a=
 \sigma_{H,n+1}^2\left( \frac{\phi_n\left(\sigma_\theta^2+\sigma_{H,n}^2 \right)+\left\langle J\right\rangle_n \alpha_n \sigma_{H,n}^2}{\sigma_{H,n}^2 \sigma_\theta^2+\sigma_{H,n}^4+\alpha_n^2 \epsilon^2} +\frac{\pi}{\sigma_\theta^2} \right).
\end{eqnarray}
Using this and Eq. \eqref{ecor3} we find $\phi_{n+1}$ as:
\begin{equation}
 \phi_{n+1}= \phi_a+\Delta\phi~\left(\mathrm{mod}\left(2 \pi \right)\right) \label{phir}
\end{equation}
with $\Delta \phi=\ell \pi+ 2 \ell \pi \beta$.

Following the same procedure we determine $\alpha_a$ from the term proportional to both $\theta$ and $J$:
\begin{eqnarray}
 \frac{2 \alpha_a}{\sigma_{H,n+1}^2}=\frac{2 \alpha_n}{\sigma_{H,n}^2}-\frac{2 \left( \sigma_{H,n}^4+ \epsilon^2 \alpha_n^2  \right)\alpha_n}{\sigma_{H,n}^2 \left(\sigma_\theta^2\sigma_{H,n}^2+\sigma_{H,n}^4+\epsilon^2\alpha_n^2\right)}\\
 \Updownarrow \nonumber \\
 \alpha_a=
 \alpha_n \sigma_{H,n+1}^2\frac{\sigma_\theta^2}{\sigma_{H,n}^2\sigma_\theta^2+\sigma_{H,n}^4+\alpha_n^2\epsilon^2}
\end{eqnarray}
which using Eq. \eqref{ecor3} gives:
\begin{equation}
 \alpha_{n+1}=\alpha_a+1=\alpha_n \sigma_{H,n+1}^2\frac{\sigma_\theta^2}{\sigma_{H,n}^2\sigma_\theta^2+\sigma_{H,n}^4+\alpha_n^2\epsilon^2}+1 \label{alp}.
\end{equation}

Finally, $\left\langle J \right\rangle$ does not change due to evolution between the standing wave pulses, Eq. \eqref{ecor3}, so $\left\langle J\right\rangle_{n+1}$ is found from the term proportional to $J$ but not $\theta$ in Eq. \eqref{neww}: 
\begin{eqnarray}
 -\frac{2 \alpha_a \phi_a}{\sigma_{H,n+1}^2}+ \frac{2 \sigma_{H,n+1}^2 \left\langle J \right\rangle_{ n+1}}{\epsilon^2}=\frac{\epsilon^2 \sigma_{H,n}^22\frac{\sigma_\theta^2}{\epsilon^2}\left(\frac{\sigma_{H,n}^2}{\epsilon^2}\left\langle J \right\rangle_{n}-\frac{\alpha_n \phi_n )}{\sigma_{H,n}^2}\right)}{ \sigma_\theta^2\sigma_{H,n}^2 +\sigma_{H,n}^4 + \epsilon^2 \alpha_n^2}\\
 \Updownarrow \nonumber \\
 \left\langle J \right\rangle_{n+1}=\frac{ \epsilon^2 \alpha_a \phi_a}{\sigma_{H,n+1}^4}+\frac{\epsilon^2 \sigma_{H,n}^2\sigma_\theta^2 \left( \frac{\sigma_{H,n}^2 \left\langle J \right\rangle_{n}}{\epsilon^2}-\frac{\alpha_n \phi_n}{\sigma_{H,n}^2}  \right)}{\sigma_{H,n+1}^2 \left(\sigma_\theta^2\sigma_{H,n}^2+\sigma_{H,n}^4+\epsilon^2\alpha_n^2\right)}. \label{JM}
\end{eqnarray}
\\
Finally, Eq. \eqref{SN} comes from integrating Eq. \eqref{neww}.

\end{widetext}

\subsection{Special case $\epsilon=0$}
When $\epsilon=0$ simple insertion into Eq. \eqref{SH} reveals that $\sigma_H$ after $N$ standing wave pulses is:
\begin{equation}
  \sigma_{H,N}^2=\frac{\sigma_\theta^2}{N}. \label{SHN}
\end{equation}
Using this and the fact that $\left\langle J\right\rangle=0$ in Eq. \eqref{phir} allows one to show
\begin{equation}
    \phi_N=\frac{\ell \pi + 2 \ell \pi \beta}{2}\left(N+1 \right)+\pi \label{phiN}.
\end{equation}

Equation \eqref{anabe} then follows from using Eq. \eqref{SHN} and \eqref{phiN} in Eq. \eqref{SN}.

\end{document}